\documentclass{aa}
\usepackage{graphicx}
\usepackage{txfonts}
\begin{document}
    
\title{Observational studies of Cepheid amplitudes. II\\
Metallicity dependence of pulsation amplitudes}

\author{L\'aszl\'o Szabados \and 
P\'eter Klagyivik}

\institute{Konkoly Observatory, H-1121 Budapest XII, Konkoly Thege \'ut 15-17.,
 Hungary}

\offprints{L. Szabados, \email{szabados@konkoly.hu}}

\date{Received  / Accepted }
                                                                                
\abstract
{
Physical and phenomenological properties (radius, luminosity, 
shape of the light curve,  etc.) of Cepheids strongly depend on 
the pulsation period, with the exception of the pulsation amplitude. 
A possible factor causing a wide range of pulsation amplitudes 
might be the different atmospheric metallicities of individual 
Cepheids.
}
{
We studied the influence exerted by the atmospheric iron content, 
[Fe/H], on the pulsational amplitude of Galactic Cepheids.}
{
We searched for correlations between the [Fe/H] value and both the 
observed amplitudes and amplitude related parameters.
}
{
The amplitude of the Cepheid pulsation slightly decreases with 
increasing iron abundance. This effect is more pronounced for the 
radial velocity variations and for the shorter pulsation periods.
The wavelength dependence of
photometric amplitudes is also found to be sensitive to the metallicity. 
Some of these effects are not consequences of differential line blanketing.
Based on the calibrations of the metallicity sensitivity relationships, 
we derived photometric iron abundance for 21 Galactic Cepheids.
The dichotomic behaviour dividing Galactic Cepheids that pulsate in the 
fundamental mode into short- and long-period groups 
at the period of 10\fd47 can be noticed in some diagrams that show
metallicity-related dependences. 
}
{
We confirm that variety in atmospheric metallicity in Cepheids
contributes to the finite range of pulsation amplitudes at a given 
period. Effects of metallicity on the amplitudes revealed from 
observational data and the occurrence of the dichotomy also derived 
from phenomenological data have to be confirmed by appropriate 
theoretical models of stellar structure and pulsation.}
{}

\titlerunning{Metallicity dependence of Cepheid amplitudes}
\authorrunning{Szabados \& Klagyivik}
                                                                                
\maketitle
                                                                                
\section{Introduction}
\label{sect_1}

Classical Cepheids are radially pulsating supergiant stars. 
This type of pulsation, maintained by the $\kappa$-mechanism 
in the outer layers of the star, is stable in a narrow range of 
the effective temperature, in the nearly vertical 
{\em instability strip} in the Hertzsprung-Russell diagram. 
The fact that this pulsation is a free radial oscillation 
of the star results in a relationship between the period, $P$, 
and luminosity, $L$, of Cepheids. 

Owing to the existence of the period-luminosity ($P$-$L$) 
relationship, Cepheids are primary distance indicators in astronomy. 
The precise calibration of this relationship has been in the 
forefront of Cepheid-related studies for decades. A recently emerged
aspect of these studies is the role of metallicity in shaping 
the $P$-$L$ relationship. In spite of thorough theoretical and 
observational investigations, the results on the metallicity
dependence are controversial (see summaries by \cite{Retal08};
\cite{M09}; \cite{FM10}).

Along with other Cepheid-related relationships, 
the $P$-$L$ relationship is roughly linear if plotted against 
$\log P$. The only exception is the period-amplitude
($P$-$A$) relationship. The wide range of possible photometric
and radial velocity amplitudes at a given pulsation period 
may be caused by various effects (pulsation mode and energy, 
companion star(s), metallicity, etc.).

Motivated by the fact that the dependence of the pulsation 
amplitude on $\log P$ is neither linear nor unique, 
we studied the $P$-$A$ relationship of Galactic Cepheids 
in Paper~I (\cite{KSz09}).
We revised the $P$-$A$ graphs for the $U$, $B$, $V$, 
$R_{\rm C}$, and $I_{\rm C}$ photometric bands.
In addition, a $P$-$A$ graph was constructed using the
observed peak-to-peak amplitudes of the pulsational radial velocity 
variations updating the much earlier diagrams (including that 
by \cite{CS84} to which we inadvertently missed to refer in Paper~I). 
One purpose of constructing new 
$P$-$A$ diagrams has been to study the possible effect of metallicity 
on the pulsational amplitude, which may contribute to the observed 
finite range of actual amplitudes at a given pulsation period.

\begin{table*}[t]
\caption{Average [Fe/H] values for various groups of Cepheids}
\begin{tabular}{lcccccccccccccc}
\hline
\noalign{\smallskip}
Sample && \multicolumn{3}{c}{Solitary Cepheids} && &\multicolumn{3}{c}{Binary
Cepheids}& &&\multicolumn{3}{c}{Total sample}\\
\noalign{\smallskip}
\cline{3-5} \cline{8-10} \cline{13-15}
 && [Fe/H] & $\sigma$ & $N$ &&& [Fe/H] & $\sigma$ & $N$ &&&[Fe/H] & $\sigma$ & $N$ \\
\noalign{\smallskip}
\hline
\noalign{\smallskip}
Fundamental mode (F) Cepheids, $\log P > 1.02$ && 
0.167 & 0.164 & 49 &&& 0.184 & 0.154 & 34 
&&& 0.174 & 0.160 & 83\\
Fundamental mode (F) Cepheids, $\log P < 1.02$ && 
0.091 & 0.158 & 114 &&& 0.103 & 0.100 & 78 
&&& 0.096 & 0.137 & 192\\
First overtone (1OT) Cepheids  && 
0.108 & 0.142 & 28 &&& 0.068 & 0.118 & 24 
&&& 0.089 & 0.133 & 52\\
\noalign{\smallskip}
\hline
\end{tabular}
\label{tab_aveFe}
\end{table*}

An influence of the chemical composition on the amplitude 
of pulsation is expected based on different patterns of the 
$P$-$A$ plots for Cepheids in our Galaxy and the Magellanic Clouds.
van Genderen (\cite{vG78}) was the first who attributed 
the differences in the largest pulsation amplitudes of
Cepheids to the different average metallicities of the respective
host galaxies. If atmospheric metallicity has an influence on 
largest possible amplitude at a given pulsation period, 
then it should exert influence on the actual amplitude 
of individual Cepheids.

Here we study the effect of the atmospheric iron content, [Fe/H],
on peak-to-peak amplitudes and amplitude-related parameters 
introduced in Paper~I for classical Cepheids of our Galaxy. 
Input values have been taken from Paper~I (where the sources 
of the amplitude and spectroscopic [Fe/H] data as well as 
the method of their homogenization is also described); 
in addition, [Fe/H] values published recently by Luck \& Lambert 
(\cite{LL11}) and Luck et~al. (\cite{Letal11}) 
have also been taken into account (after applying 
the same correction as for the [Fe/H] data tabulated in Paper~I).
The present sample consists of 329 Galactic Cepheids with 
known spectroscopic [Fe/H] value.

This paper is organized as follows. Section~\ref{sect_per} 
deals with the period dependence of the iron content of 
Galactic Cepheids. Correlations between the [Fe/H] value and 
various amplitudes and amplitude-related parameters 
are presented in Sect.~\ref{sect_amp}. Consequences of our results
are discussed in Sect.~\ref{sect_disc}, while the conclusions 
are drawn in Sect.~\ref{sect_concl}.

\section{Period dependence of [Fe/H]}
\label{sect_per}

\subsection{Period - {\rm [Fe/H]} relationship}
\label{ssect_permetrel}

Long-period Cepheids, which are more luminous (thus more massive)
than their shorter period siblings, evolve more rapidly. 
As a consequence, longer period Cepheids are also younger, which 
is expressed by the period-age relationship (see, e.g.,  
\cite{E03}). Owing to the continuous enrichment of the 
interstellar matter with heavy elements, the metal abundance 
of the younger stellar population is higher. 

\begin{figure}[hb!]
   \centering
   \includegraphics[width=80mm]{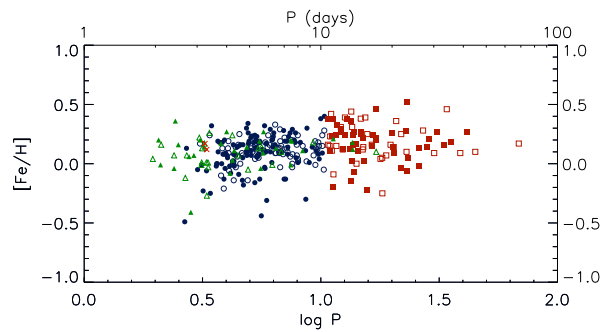}
   \caption{Dependence of [Fe/H] on the pulsation period. 
Circles represent F-mode Cepheids pulsating 
with short period ($P < 10\fd47$), squares denote their 
long-period counterparts ($P > 10\fd47$), while triangles
correspond to 1OT pulsators. Empty and filled symbols 
are used for distinguishing Cepheids with and without known 
companions, respectively.}
   \label{fig:period}
   \end{figure}

Figure~\ref{fig:period} shows the [Fe/H] value as a function 
of $\log P$. In all figures throughout this paper, Cepheids 
pulsating in the fundamental (F) mode are denoted by (blue) circles 
if $P < 10\fd47$ and (red) squares for $P > 10\fd47$, while (green) 
triangles correspond to first overtone (1OT) pulsators. The 
replacement of the conventional period limit (10 d) separating 
short- and long-period Cepheids by $10\fd47$ was justified in 
Paper~I. Empty and filled symbols are used for distinguishing 
Cepheids with and without known companions, respectively. In most 
of the subsequent figures, however, Cepheids with known companion(s) 
were disregarded because companion stars may falsify 
amplitudes.
According to various pieces of evidence
(see http://www.konkoly.hu/CEP/intro.html),
137 Cepheids (of 329) have companion(s). This high frequency of 
occurrence of binaries (or multiple systems) in this Cepheid sample 
agrees with the statistics presented by Szabados (\cite{Sz03}).
 
As expected, the average [Fe/H] value slightly increases towards 
longer pulsation periods in Fig.~\ref{fig:period}, although there is 
a wide range of observed iron abundances at any period. This last 
observation is a consequence of the dependence of the chemical composition 
of stars on Galactic position (see Sect.~\ref{ssect_galdistr}).

The average value of the iron abundance of separate groups
of Cepheids (F mode, short and long period; 1OT) 
is summarized in Table~\ref{tab_aveFe}, separately for
`solitary' Cepheids (i.e., for Cepheids without known companions),
Cepheids with companions, and the whole sample. 

The period dependence of the iron abundance was determined by
the least-squares method. The line expressing the increase in 
the iron abundance with the pulsation period is
\begin{equation}
{\rm [Fe/H]} = -(0.012 \pm 0.027) + (0.149 \pm 0.030) \times \log P
\label{perioddep}
\end{equation}
\noindent based on all 329 Cepheids in the sample.
Freedman \& Madore (\cite{FM11}) could
not find any period dependence of [Fe/H] values for a sample of 
22 Cepheids in the Large Magellanic Cloud. This contradicting result 
can be explained by the lack of short-period Cepheids in their sample.

This dependence is robust, it is not an artefact caused by
outliers (i.e., most metal deficient) Cepheids. 
Omitting the six most deviating [Fe/H] values ([Fe/H$]< -0.25$,
the fit becomes
\begin{equation}
{\rm [Fe/H]} = 0.017 \pm 0.023) + (0.125 \pm 0.027) \times \log P \ \ .
\label{perioddeprobust}
\end{equation}


\subsection{Cepheid desert in the 8-10 day period interval}
\label{ssect_misscep}

\begin{figure}[b!]
\begin{center}
\includegraphics[width=80mm]{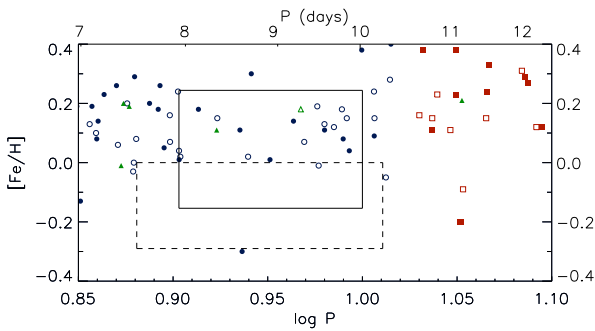}
\end{center}
\caption{Distribution of [Fe/H] values near the `zone of 
avoidance'. Cepheids with known companions are also plotted.
The meaning of the symbols is the same as in Fig.~\ref{fig:period}.
The theoretically predicted region of the Galactic Cepheid desert
is represented by the solid line box, while the observed 
desert region is marked as the dashed line box.}
\label{fig:zoom}
\end{figure}

The $P$-$A$ graphs show a prominent dip in the pulsation amplitude 
at $\log P = 0.98$ for Galactic Cepheids (Paper~I). At slightly
shorter pulsation periods, there is a deficiency of Cepheids 
in the interval of $ 0.930 < \log P < 0.965$. This `zone of avoidance' 
was first mentioned by van~Genderen (\cite{vG70}). 
A similar deficit is present in the period distribution of 
Cepheids in M31 but no such feature appears in the $P$-$A$ plots
for Magellanic Cepheids.

Nonlinear pulsation calculations were able to successfully reproduce 
this behaviour in the observed number distribution. Buchler et~al.
(\cite{BGP97}) found that the fundamental mode limit cycles are 
unstable in the period domain $8-10$ d for Cepheids with metallicity
in the range $Z = 0.014-0.035$. The corresponding stars pulsate in the 
first overtone with a period of $P_1 \approx 0.7 P_0$, which causes a 
relative excess of 1OT Cepheids in the $5\fd6-7\fd0$ period interval. 
This concerns our Galaxy and M31, as well. Magellanic Clouds 
are of lower metallicity, therefore F-mode Cepheid pulsation 
is maintained in them in the period interval of $8-10$ d. 
According to this theoretical prediction, Galactic Cepheids within 
the period interval of $8-10$ d may contain a much greater or lesser 
amount of heavy elements in their atmosphere than the Galactic average.  

Using the transformation formula between the heavy element abundance, 
$Z$, and the iron content, [Fe/H], given by Caputo et~al. (\cite{Cetal01})
\begin{equation}
{\log Z } = {\rm [Fe/H]} - 1.7~~,
\label{metaliron}
\end{equation}
the $0.014 < Z < 0.035$ interval of metallicity corresponds to a
$-0.154 < [{\rm Fe/H}] < 0.244$ range of iron abundance.

The relevant part of Fig.~\ref{fig:period} is shown separately
in Fig.~\ref{fig:zoom}. The box drawn with a solid line represents
the theoretically obtained `zone of avoidance' as calculated by
Buchler et~al. (\cite{BGP97}), while the borders of the observed 
Galactic Cepheid desert are marked with the box drawn with a dashed 
line. Our sample shows a deficiency of the F-mode Cepheids in the
$ 0.88 < \log P < 1.01$ interval for the metallicity 
range of $ 0.00 > {\rm [Fe/H]} > -0.29$.

The region of the theoretical box only partially overlaps with the
actual desert region of Cepheids that pulsate in the F-mode. The observed 
limits of the `zone of avoidance' imply a wider period interval in 
both directions and suggest that the mechanism computed by 
Buchler et~al. (\cite{BGP97}) is only valid for the $Z = 0.010-0.020$
range. A revision of the theoretical picture is suggested.

\subsection{Galactic distribution of Cepheid metallicities}
\label{ssect_galdistr}

\begin{table*}
\caption{Coefficients of the linear fit to the amplitude versus [Fe/H]
relationships (Eq.~\ref{fe-amp}) based on solitary Cepheids.
$N$ is the number of Cepheids in the given sample, $r$ is the
correlation coefficient.}
\begin{tabular}{lc@{\hskip2mm}cccccccccccccc}
\hline
\noalign{\smallskip}
&  \multicolumn{4}{c}{$A_{V_{\rm RAD}}$} && \multicolumn{4}{c}{$A_B$} &&\multicolumn{4}{c}{$A_R$}\\
\cline{2-5} \cline{7-10} \cline{12-15}
\noalign{\smallskip}
Sample& $a$ & $b$  & $N$ &$r$&& $a$ & $b$ & $N$ &$r$&& $a$ & $b$ & $N$ & $r$\\
& ($\sigma_a$) & ($\sigma_b$) &&&& ($\sigma_a$) & ($\sigma_b$) &&&& ($\sigma_a$) & ($\sigma_b$) && \\
\noalign{\smallskip}
\hline
\noalign{\smallskip}
F Cepheids, $\log P > 1.02$  & 
53.39 & $-$17.57 & 31 &$-$0.239&& 1.620 &
$-$0.254 & 44 & $-$0.154&& 0.817 & $-$0.009 &
46 & $-$0.011\\
& (0.85) & (7.74) &&&& (0.066) & (0.255) &&&&
(0.029) & (0.114) &&\\
F Cepheids, $\log P < 1.02$  & 
37.01 & $-$24.58 & 58 & $-$0.515 && 1.054 & 
$-$0.273 & 105 &$-$0.237&& 0.571 &$-$0.162 & 
108 & $-$0.239\\
& (2.36) & (5.47) &&&& (0.021) & (0.110) &&&& 
(0.012) & (0.063) &&\\
1OT Cepheids  & 
17.55 & $-$13.00 & 24 & $-$0.340 && 0.500 &
$-$0.339 & 28
&$-$0.398&&0.277 & $-$0.170 & 27 & $-$0.347 \\
& (1.59) & (7.87) &&&& (0.022) & (0.127) &&&& (0.015) &
(0.078) &&\\
\noalign{\smallskip}
\hline
\end{tabular}
\label{tab_Amp_Fe}
\end{table*}

\begin{figure}[h!]
\begin{center}
\includegraphics[width=80mm]{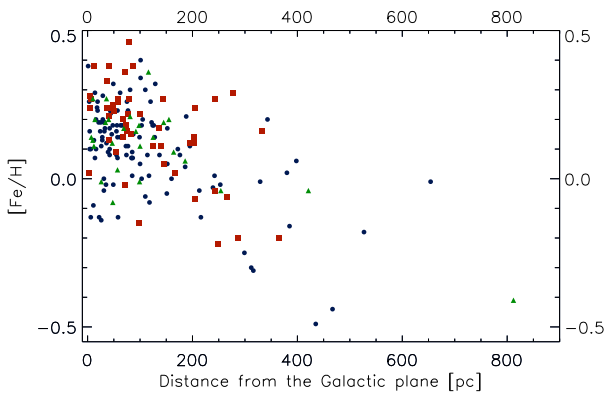}
\end{center}
\caption{Distribution of [Fe/H] as a function of the distance from 
the Galactic plane. The meaning of the symbols is the same as 
in Fig.~\ref{fig:period}.}
\label{fig:gallat}
\end{figure}

The metal abundance of stars depends on the location in the host
galaxy. The dependence of [Fe/H] on galactic longitude in Cepheids
has been studied by Andrievsky et~al. (\cite{Aetal04}) -- who also 
discussed the radial metallicity gradient observed in our Galaxy
-- and by Klagyivik \& Szabados (\cite{KSz07}). 
Here we only study the distribution of [Fe/H] values in the direction 
perpendicular to the Galactic plane. The distance from the Galactic 
plane was calculated for each Cepheid from the galactic latitude and 
the distance (taken from the Cepheid data base maintained at the 
David Dunlap Observatory -- see \cite{FBES95}).
The results are shown in Fig.~\ref{fig:gallat}. Here and in the
subsequent figures Cepheids with known companions have been
disregarded. 

The diagram shows a trend that Cepheids more distant from 
the Galactic plane are more deficient in iron. 
In this case, however, we are not facing an age effect.
Classical Cepheids belong to the young stellar population, 
therefore the pattern seen in Fig.~\ref{fig:gallat} does not imply
a dependence of [Fe/H] on the distance from the Galactic plane
for the sample Cepheids. Instead, it can be interpreted as an
artefact of merging Cepheids of all galactic longitudes in a
single diagram. Most metal-poor Cepheids are situated in the
outer spiral arm (constellations Aur, Cam, Cas, Per) where the
disk extends farther perpendicularly to the Galactic plane than 
at shorter galactocentric distances (\cite{BM98}). 
The warp of the Galactic plane also
contributes to the pattern seen in Fig.~\ref{fig:gallat}.
From our viewpoint, the Galactic disk turns up to the north
in the direction of Cygnus at $\ell \simeq 90^o$, and
south in the direction of Vela at $\ell \simeq 270^o$
(\cite{P09}).

\section{Effects of the iron content on the pulsation}
\label{sect_amp}

The pulsation amplitude of the Cepheids depends on several 
factors (period, presence of companions, mode and energy of 
pulsation -- these latter two factors are closely related 
to the location of the star within the instability region),
therefore it is not easy to distinguish the effect of metallicity.

\subsection{{\rm [Fe/H]} dependence of photometric amplitudes}
\label{ssect_amp}

It is a realistic assumption that metallicity has an influence 
on the amplitude of pulsation. Using the upper envelope of the 
$P$-$A$ relationship derived in Paper~I, no reasonable correlation 
could be revealed between the `amplitude defect' (at a given 
period), $\Delta A = A_{{\lambda}_{\rm max}} - A_{\lambda}$ 
and the iron content in any photometric band, nor for the 
radial velocity amplitudes.

\begin{figure}[htb!]
   \centering
   \includegraphics[width=8cm]{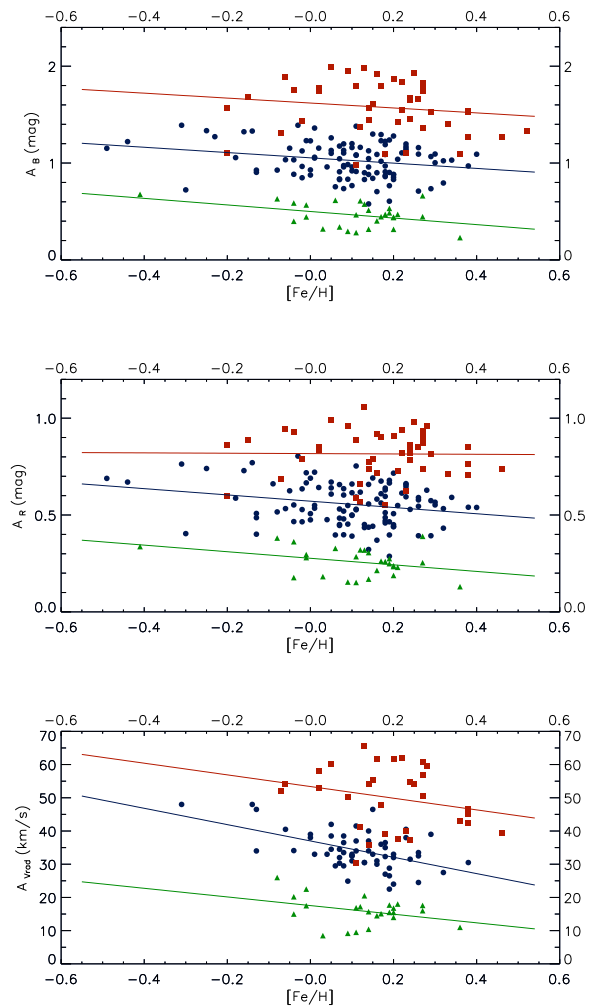}
   \caption{Dependence of photometric $B$ (upper panel),
   $R_C$ (middle panel), and radial velocity amplitudes 
(lower panel) on the atmospheric iron content. The meaning of 
the symbols is the same as in Fig.~\ref{fig:period}.
The linear least-squares fit to each subsample is also plotted.}
   \label{fig:AFe}
   \end{figure}

It turned out, however, that the amplitudes themselves show a
slight metallicity dependence if the Cepheids are properly grouped. 
Our findings are visualised in Fig.~\ref{fig:AFe} for the 
photometric $B$ (upper panel), photometric $R_{\rm C}$ (middle panel), 
and the radial velocity amplitudes (lower panel). 
The metallicity effect is conspicuous for Cepheids pulsating 
in the 1OT and for short-period F-Cepheids, in the sense that 
{\em more metal-deficient Cepheids pulsate with a larger amplitude}. 
The correlation is most pronounced between the radial velocity
amplitude and the iron content.
There are several deviating points among long-period Cepheids:
the low amplitude of the longest period Cepheids ($\ell$~Car, RY~Vel)
is a known empirical fact (see Paper~I) and some F-Cepheids whose 
period is near the value of the dichotomy (e.g., VX~Per) mimic
the behaviour of Cepheids with periods shorter than the limit of 
$10\fd47$). 

Table~\ref{tab_Amp_Fe} summarizes the results of linear 
least-squares fits. Instead of the widely used ordinary 
least-squares regression of the dependent variable $y$ against the 
independent variable $x$, or OLS($y|x$), we used the OLS bisector 
method (\cite{Ietal90}) in these cases and all
subsequent fits in this paper.
This regression line treats the variables symmetrically instead 
of the OLS($y|x$), which minimizes the sum of the squares of the 
$y$ residuals. 
The coefficients $a$ and $b$ in Table~\ref{tab_Amp_Fe}
correspond to the following formula:
\begin{equation}
A_X = a + b \times \rm [Fe/H]~~,
\label{fe-amp}
\end{equation}
\noindent where the subscript $X$ refers to the relevant band.
ER~Aur has been omitted from the figures and the fits because
its photometric amplitude turned out to be indefinite on closer
inspection.

Figure~\ref{fig:AFe} also confirms that the dichotomy between 
short- and long-period Cepheids occurs at $P = 10\fd47$. Dividing 
fundamental-mode Cepheids into two groups at the conventional value 
of $P = 10\fd0$, plots corresponding to Fig.~\ref{fig:AFe} resulted 
in a less clear picture with more outliers.

\subsection{Dependence of amplitude parameters on {\rm [Fe/H]}}
\label{ssect_q}

The importance of the amplitude ratio $A_{V_{\rm RAD}}/A_B$
(referred to as the $q$ parameter) was discussed in Paper~I. 
It is sensitive to both the oscillation mode and the
presence of a companion to the Cepheid. Unfortunately, the two 
effects cannot be separated from each other, which diminishes the 
diagnostic value of the $q$ parameter. An additional complication
is that $q$ also depends on the metallicity of the pulsating star,
as shown below.

\begin{figure}[htb!]
   \centering
   \includegraphics[width=7.5cm]{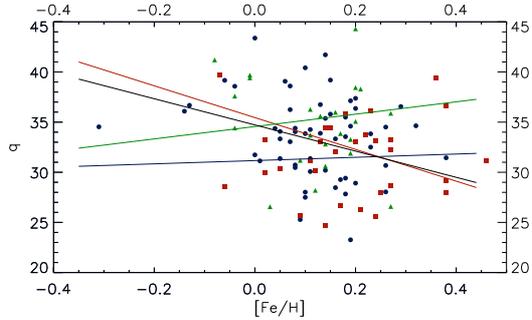}
   \caption{Metallicity dependence of the $q$ amplitude ratio. 
The various fits are explained in the text.}
   \label{fig:q}
   \end{figure}

The relationship between $q$ and [Fe/H] is shown 
in Fig.~\ref{fig:q}, involving only solitary Cepheids.
The meaning of the symbols is the same as in Fig.~\ref{fig:period}.
Linear least-squares fits have been applied to each group
separately, and the results are summarized in Table~\ref{tab_AR_Fe}.
Coefficients $a$ and $b$ correspond to the formula
\begin{equation}
q = a + b \times \rm [Fe/H]~~. 
\label{fe-ar}
\end{equation}
The last two columns of Table~\ref{tab_AR_Fe} give the number
of data points, $N$, involved in the linear regression analysis 
and the correlation coefficient, $r$.
The fits are also shown in Fig.~\ref{fig:q}: 
the red linear section is for the long-period F 
pulsators, the blue section for their short-period counterparts,
the black line denotes the fit to the sample of all F Cepheids, 
while the green section represents the fit for the 1OT Cepheids.

\begin{table}[b]
\caption{Coefficients of the linear fit to the $q$ versus [Fe/H]
relationship (Eq.~\ref{fe-ar}) based on solitary Cepheids}
\begin{tabular}{lc@{\hskip2mm}c@{\hskip2mm}c@{\hskip1mm}c@{\hskip2mm}c@{\hskip2mm}c}
\hline
\noalign{\smallskip}
Sample& $a$ & $\sigma_a$ & $b$ & $\sigma_b$ & $N$ & $r$ \\
\noalign{\smallskip}
\hline
\noalign{\smallskip}
F Cepheids, $\log P > 1.02$    & 31.17 & 1.97 & $-$15.85 
& 6.05 & 30 & 0.017\\
F Cepheids, $\log P < 1.02$    & 35.45 & 0.88 & 1.63 
& 9.34 & 56 & $-$0.236\\
All F mode Cepheids     & 34.72 & 0.89 & $-$13.06 
& 5.51 & 86 & $-$0.204\\
1OT Cepheids       & 34.56 & 3.15 & 6.16 & 25.27 
& 24 & 0.065 \\
\noalign{\smallskip}
\hline
\end{tabular}
\label{tab_AR_Fe}
\end{table}

In Paper~I we quantified the wavelength dependence of the photometric
amplitudes with two numerical parameters. Plotting the amplitude of 
the light variations measured in the $U$, $B$, $V$, and $R_C$ bands
normalized to the amplitude observed in $B$ band against 
$1/\lambda$, the distribution of the points is roughly linear 
for a given Cepheid (\cite{F79}). The slope of the straight 
line fitted to these points has been defined as the $m$ parameter 
for individual stars (Paper~I).

   \begin{figure}[htb!]
   \centering
   \includegraphics[width=7.5cm]{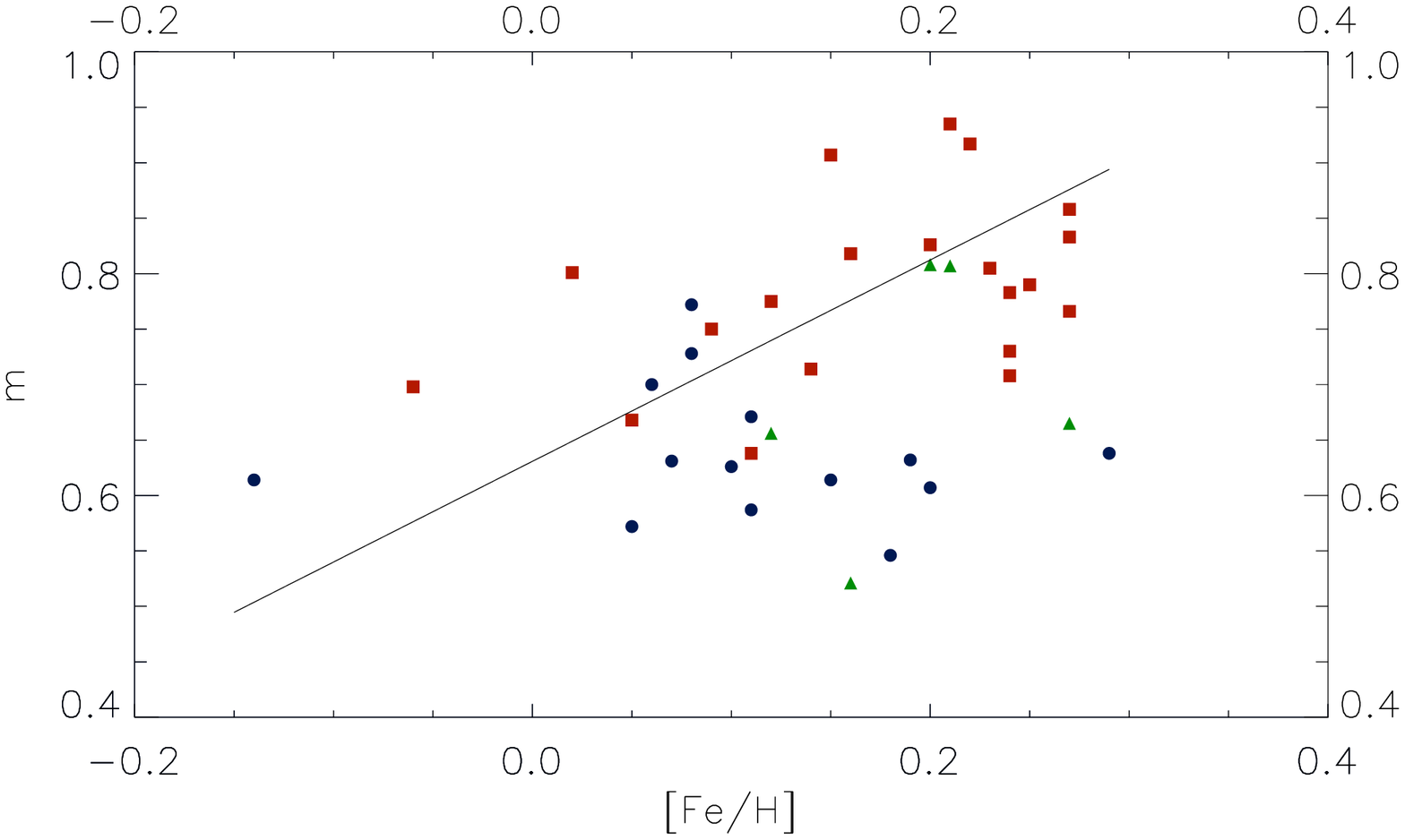}
   \includegraphics[width=7.5cm]{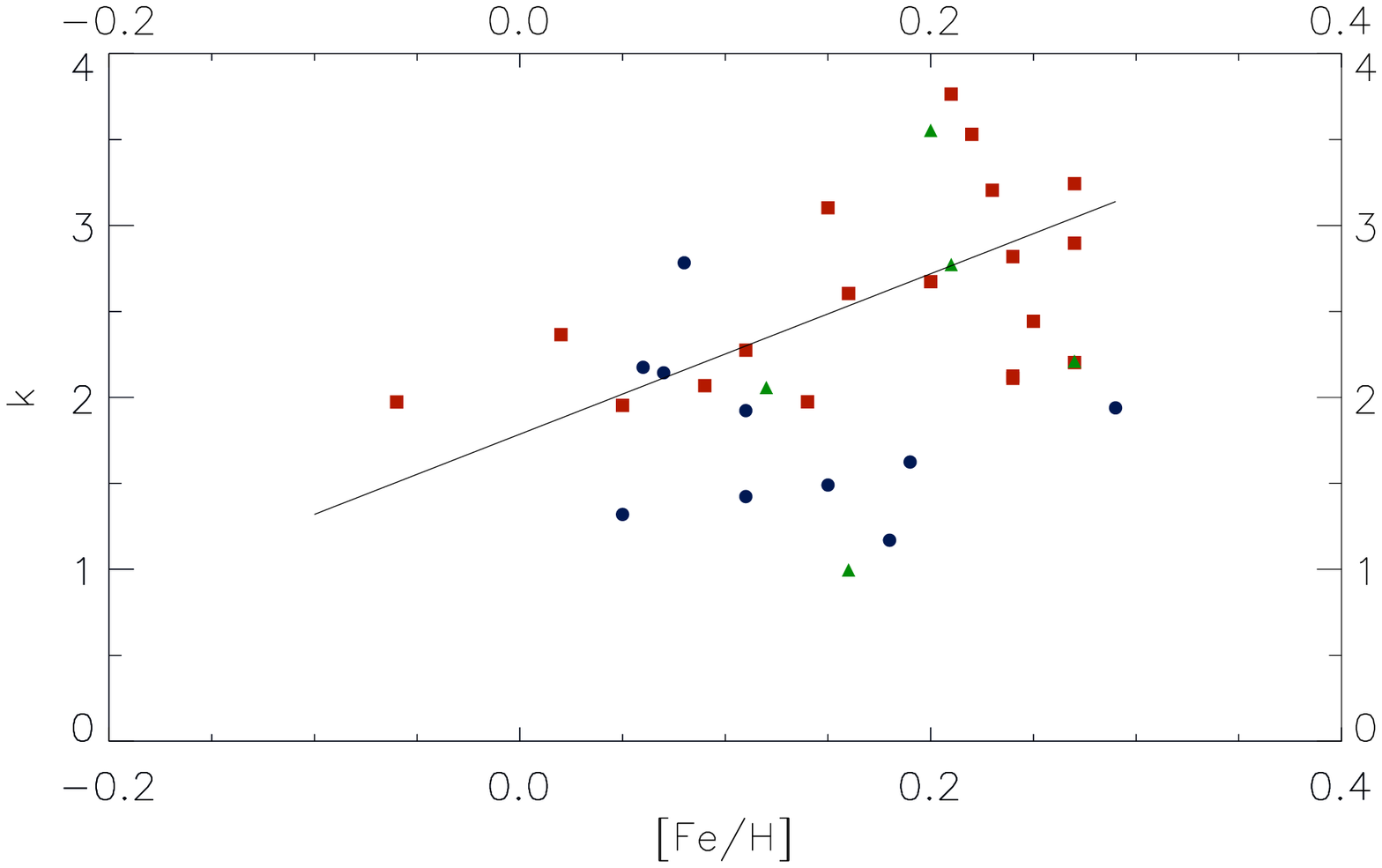}
   \caption{Metallicity dependence of the parameters $m$ 
   (upper panel) and $k$ (lower panel). The meaning of the 
   symbols is the same as for Fig.~\ref{fig:period}. 
   The linear least-squares fit to the sample of 
   long-period Cepheids is also shown.}
   \label{fig:mk}
   \end{figure}

   \begin{figure}[htb!]
   \centering
   \includegraphics[width=7.5cm]{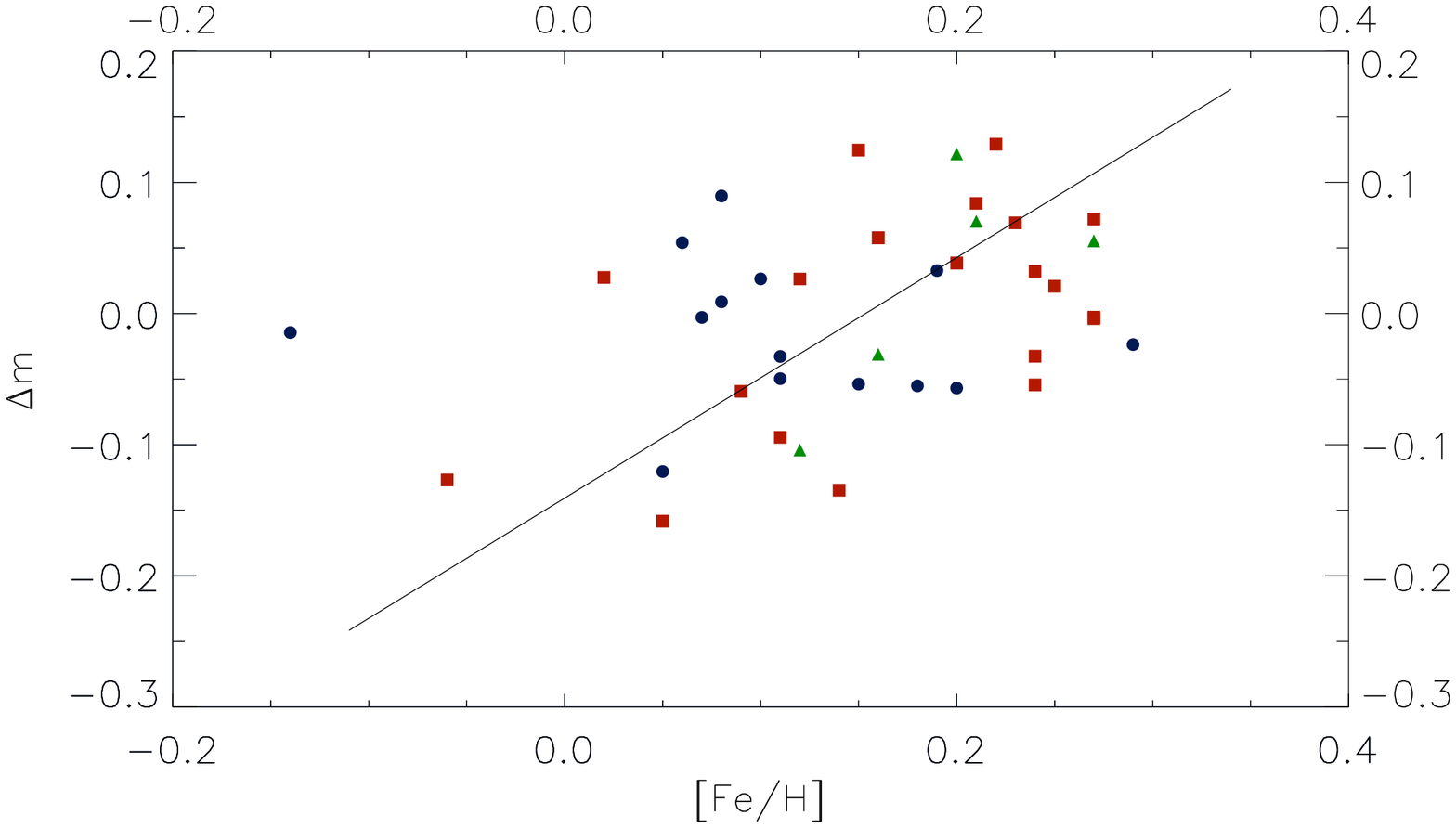}
   \includegraphics[width=7.5cm]{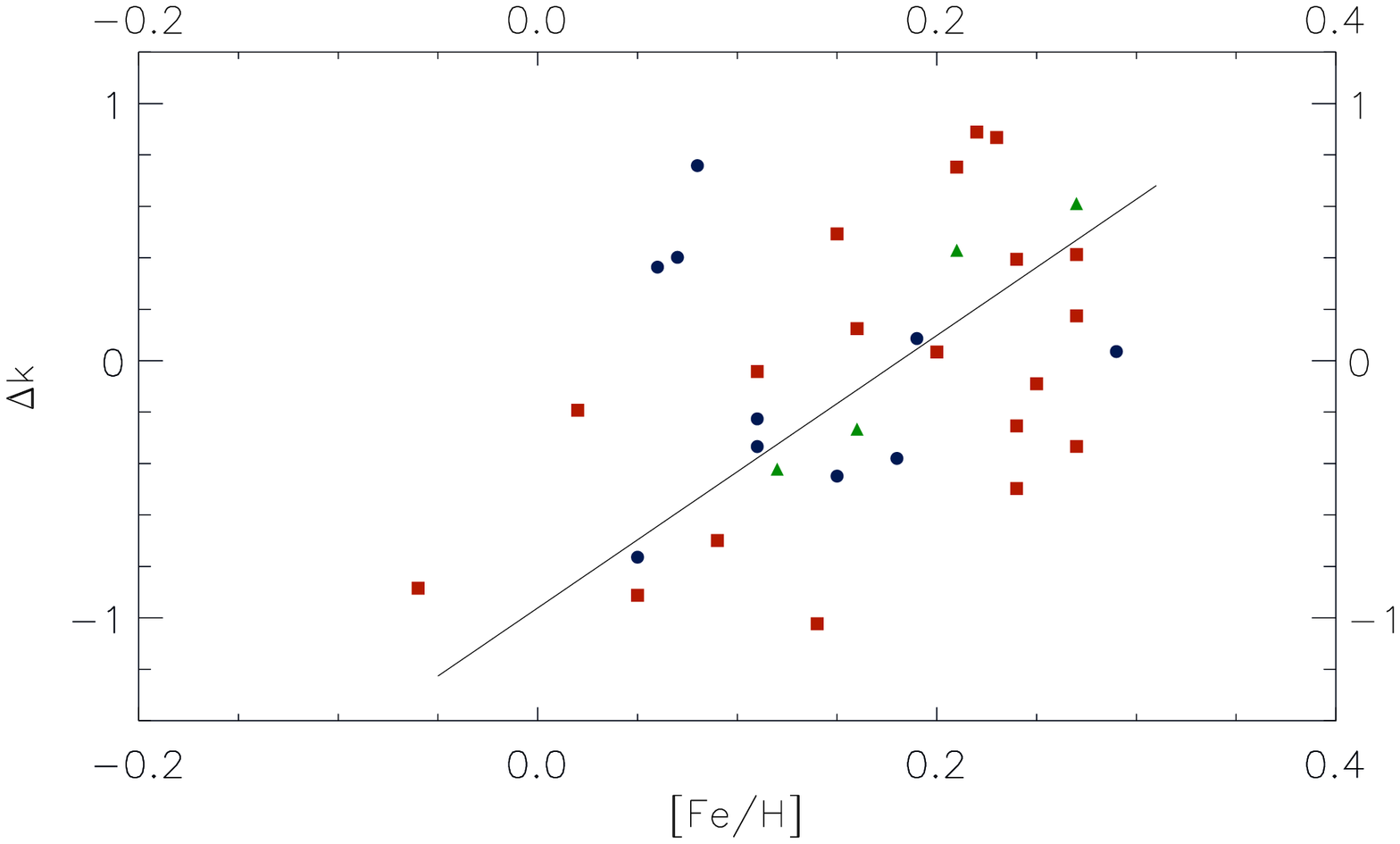}
   \caption{Metallicity dependence of $\Delta m$ and $\Delta k$ 
   parameters and the linear least-squares fit for the solitary,
   long-period Cepheids. The meaning of the symbols is the same 
   as for Fig.~\ref{fig:period}.}
   \label{fig:deltamk}
   \end{figure}

The top panel of Fig.~\ref{fig:mk} shows the relation between the 
iron content and the $m$ parameter for F-mode solitary Cepheids.
(Data for four 1OT Cepheids are also plotted.) In the case of 
long-period Cepheids, metallicity has a noticeable effect.
The higher the [Fe/H], the larger the $m$ parameter, i.e., 
the amplitude decreases stronger with increasing wavelength. 
The linear fit to the points representing 20 Cepheids with 
$P > 10\fd47$ 
(also shown in the upper panel of Fig.~\ref{fig:mk}) is
\begin{equation}
m = 0.631 (\pm 0.033) + 0.908(\pm 0.142) \times \rm [Fe/H]~~.
\label{m-slope}
\end{equation}
\noindent The correlation coefficient of the linear regression is 
$r=0.416$.

The other parameter, $k$, characterizes the wavelength dependence
of photometric amplitudes in five ($U$, $B$, $V$, $R_C$, and $I_C$)
photometric bands. Approximating the amplitude-wavelength graph 
with an exponential curve of the form
\begin{equation}
{(\rm Amplitude})_\lambda = c_1 + c_2\times {\lambda}^{-k}
\label{eq-k}
\end{equation}
serves as the definition of the $k$ parameter introduced in Paper~I.

The dependence of the $k$ parameter on the atmospheric iron content
is shown in the bottom panel of Fig.~\ref{fig:mk}. The linear section
shown there corresponds to the linear least-squares fit to the data
of 19 long-period ($P > 10\fd47$) Cepheids:
\begin{equation}
k = 1.786 (\pm 0.148) + 4.665 (\pm 1.010) \times \rm [Fe/H]~~.
\label{k-slope}
\end{equation}
\noindent The correlation coefficient is $r= 0.470$.

The effect of metallicity on the wavelength dependence of 
the pulsation amplitude of the 
long-period Cepheids is again clearly seen. 

Both $m$ and $k$ parameters depend on the pulsation period, as well
(Paper~I). To quantify the real [Fe/H] dependence of these parameters,
the period dependence has to be removed. Let $\Delta m$ denote the 
difference between the actual $m$ value of a Cepheid and the mean 
$m$ calculated for the given pulsation period from Eq.~12 in Paper~I. 
Likewise, $\Delta k$ is introduced for characterizing the deviation 
of $k$ values of individual Cepheids from the ridge line given by 
Eq.~14 in Paper~I.

The $\Delta m$ and $\Delta k$ values as a function of [Fe/H] are 
plotted in Fig.~\ref{fig:deltamk}. Linear least-squares fits to the 
data for the long-period Cepheids resulted in the following equations:
\begin{equation}
\Delta m = -0.150 (\pm 0.032) + 0.923 (\pm 0.143) \times \rm [Fe/H]
\label{deltam-slope}
\end{equation}
\noindent based on 20 data points ($r=0.500$) and
\begin{equation}
\Delta k = -0.961 (\pm 0.193) + 5.297 (\pm 1.017) \times \rm [Fe/H]
\label{deltak-slope}
\end{equation}
\noindent from 19 data points ($r=0.540$). 
The subsamples of short-period Cepheids and s-Cepheids (though they 
are plotted in Fig.~\ref{fig:deltamk}) span a rather narrow 
metallicity interval and include too few stars to obtain meaningful 
results of this type. 

Figure~\ref{fig:deltamk} shows that both $\Delta m$ and 
$\Delta k$ increase with increasing iron abundance and the scatter 
is narrower than for the plots in Fig.~\ref{fig:mk}.

\subsection{{\rm [Fe/H]} values derived from the amplitudes}
\label{ssect_ampparmet}

The strength of the metallic lines in stellar spectra has
an influence on the brightness observable in particular
in blue and violet photometric bands. For intermediate-band 
photometries, even calibrations exist between intrinsic 
(i.e., reddening corrected) colour indices 
and the [Fe/H] value. Eggen (\cite{E85}) determined the 
metallicity of about a hundred Cepheids observed in the  
Str\"omgren system. The broad-band Washington system has been
elaborated specifically for determining atmospheric abundances
of stars whose temperature is similar to that of Cepheids 
(\cite{C76}). Metal abundances derived from photometric data 
obtained in this system are uncertain by a factor of two. Colour 
indices of the broad-band Johnson system (in which the overwhelming 
majority of Cepheid photometries has been carried out) are not 
ideal for abundance studies, but colour indices of this $UBV$ system 
were also calibrated in terms of [Fe/H] via the line blanketing effect 
(see \cite{Ketal11} and references therein). The drawback of the 
applications of these calibrations to photometric data on Cepheids 
is the uncertainty in the colour excess of individual Cepheids.

\begin{table}[!t]
\caption{The newly determined photometric iron abundances. 
See the explanation in the text.}
\begin{tabular}{l@{\hspace{4mm}}c@{\hspace{4mm}}rrr}
\noalign{\smallskip}
\hline
\noalign{\smallskip}
Cepheid & $P$ & [Fe/H]$_{\rm sp}$ & [Fe/H]$_{\Delta m}$ & [Fe/H]$_{\Delta k}$ \\
 & (d) &&&\\
\noalign{\smallskip}
\hline
\noalign{\smallskip}
   SZ Aql & $17.139$ & $0.22$ & $0.27$ & $0.35$ \\
   TT Aql & $13.755$ & $0.27$ & $0.20$ & $0.23$ \\
   XY Car & $12.439$ & $0.12$ & $0.15$ &   ---  \\[0.8ex]
   XZ Car & $16.650$ & $0.24$ & $0.05$ & $0.01$ \\
   RW Cas & $14.790$ & $0.27$ & $0.11$ & $0.05$ \\
    X Cyg & $16.386$ & $0.15$ & $0.26$ & $0.25$ \\[0.8ex]
   TX Cyg & $14.710$ & $0.25$ & $0.14$ & $0.11$ \\
   CD Cyg & $17.076$ & $0.20$ & $0.16$ & $0.14$ \\
 V609 Cyg & $31.080$ & $0.27$ & $0.11$ & $0.18$ \\[0.8ex]
   SV Mon & $15.233$ & $0.02$ & $0.15$ & $0.09$ \\
   UU Mus & $11.636$ & $0.24$ & $0.08$ & $0.07$ \\
    U Nor & $12.655$ & $0.24$ & $0.15$ & $0.23$ \\[0.8ex]
   VX Per & $10.889$ & $0.11$ & $0.01$ & $0.12$ \\
   VZ Pup & $23.163$ &$-0.06$ & $-0.03$ & $-0.08$ \\
   BN Pup & $13.673$ & $0.16$ & $0.18$ & $0.16$ \\[0.8ex]
   GY Sge & $51.612$ &   ---  & $0.21$ & $0.24$ \\
   KQ Sco & $28.699$ & $0.21$ & $0.22$ & $0.31$ \\
   RY Vel & $28.134$ & $0.14$ &$-0.04$ & $-0.11$ \\[0.8ex]
   RZ Vel & $20.411$ & $0.09$ & $0.05$ & $-0.03$ \\
   SW Vel & $23.436$ & $0.05$ &$-0.07$ & $-0.09$ \\
   DR Vel & $11.199$ & $0.23$ & $0.20$ & $0.34$ \\
\noalign{\smallskip}
\hline
\noalign{\smallskip}
\end{tabular}
\label{fotofeperh}
\end{table}

   \begin{figure}[t!]
   \centering
   \includegraphics[width=7.5cm]{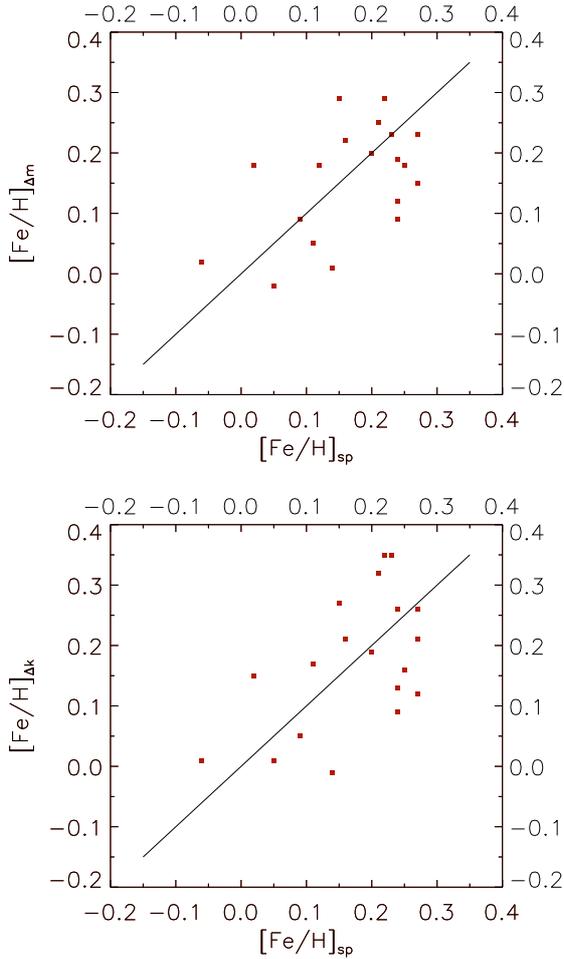}
   \caption{Correlations between photometric and spectroscopic
   [Fe/H] values. Spectroscopic [Fe/H] is indicated along the 
   abscissa. The various photometric iron abundances are 
   explained in the text. The 1:1 correspondence 
   is shown by a linear section in each panel.}
   \label{fig:feperhfeperh}
   \end{figure}

The metallicity dependence of the amplitude parameters discussed
in Sect.~\ref{ssect_q} can be used for determining the [Fe/H] of 
individual long-period Cepheids, based on Eqs.~\ref{deltam-slope} 
and \ref{deltak-slope}, if peak-to-peak photometric amplitudes 
are known, and Eq.~\ref{fe-ar}, if the amplitude of the pulsational 
radial velocity variation and the photometric $B$ amplitude are known.
We refer to these [Fe/H] values as photometric iron abundances.

Photometric and spectroscopic [Fe/H] values are compared 
in Fig.~\ref{fig:feperhfeperh} using those `solitary' Cepheids 
for which both kinds of iron abundance is known.
The vertical axis shows\\
\noindent [Fe/H]$_{\Delta m}$ (upper panel) as derived from the 
metallicity dependence of the $\Delta m$ parameter based on 
Eq.~\ref{deltam-slope},
the sample consists of 20 F Cepheids;\\ 
\noindent [Fe/H]$_{\Delta k}$ (lower panel) as derived from the 
metallicity dependence of the $\Delta k$ parameter based on 
Eq.~\ref{deltak-slope}, the sample consists of 19 F Cepheids.

Figure~\ref{fig:feperhfeperh} shows that from the 
$\Delta k$ and $\Delta m$ parameters one can 
determine an approximate [Fe/H] value for long-period Cepheids
with an uncertainty of $\pm 0.15$ dex. 
The newly determined photometric [Fe/H] values for 21 long-period
Cepheids are listed in Table~\ref{fotofeperh}, whose columns 
contain the following data: name of the Cepheid;
pulsation period (rounded to 3 decimal figures);
[Fe/H]$_{\rm sp}$, spectroscopic iron abundance taken from 
Table~1 in Paper~I or from the more recent lists by Luck \& Lambert
(\cite{LL11}) and Luck et~al. (\cite{Letal11}); 
[Fe/H]$_{\Delta m}$; [Fe/H]$_{\Delta k}$, respectively.

For the long-period Cepheid GY~Sge (0.22) this is the first 
metallicity estimation (with an estimated uncertainty of $\pm 0.04$ dex).


\section{Discussion}
\label{sect_disc}

According to our result, photometric amplitudes become larger 
with decreasing iron abundance. This is in contrast to the 
conclusion drawn by Paczy\'nski \& Pindor (\cite{PP00}), 
who compared the photometric amplitudes of long-period Cepheids 
in the two Magellanic Clouds, and found that the higher the 
metal content of the host galaxy, the larger the amplitude. 
Their study was not based on the metallicity of individual
Cepheids, instead, Paczy\'nski \& Pindor compared the median 
amplitudes and assigned the mean metallicity of the host galaxy 
to the respective amplitude value.
That method is an oversimplification because the metallicity
of individual Cepheids is a function of the location within
the host galaxy, and the sample investigated by Paczy\'nski 
\& Pindor was too small (35 Cepheids in the SMC, 33 in the
LMC, and 42 in the Milky Way Galaxy) to be representative 
for the whole galaxy. In addition, blending by unresolved
companions might also bias the photometric amplitudes 
derived for Magellanic Cepheids.

An alternative explanation for the significant difference
between the mean amplitudes of Cepheids in the three galactic 
systems can be a difference in the helium content in the
stars whose observational verification is beyond the
current spectroscopic capabilities. The location of the
instability region in the Hertzsprung-Russell diagram also
depends on the helium abundance (\cite{BIT77};
\cite{Metal05}), 
and the differing effective temperature may also alter 
the pulsational properties of Cepheids.

Based on extensive spectroscopic investigations of classical 
Cepheids, Luck \& Andrievsky (\cite{LA04}), Andrievsky et~al. 
(\cite{ALK06}), Kovtyukh et~al. (\cite{KABL06}), and Luck et~al. 
(\cite{LAFK08}) pointed out that there are no observable 
changes in the chemical abundance during the pulsational cycle.
Variations in the effective temperature during the pulsation,
however, result in variable strength of the metallic lines in spectra
of Cepheids because line blanketing is stronger during the phases
of lower temperature which in turn results in apparently larger photometric
amplitudes for more metal-deficient Cepheids, especially in the
shorter wavelength $B$ and $U$ bands. This effect is a plausible
explanation for the observed dependence of the photometric amplitudes 
on the iron content shown in Fig.~\ref{fig:AFe}.

Differential line blanketing, however, does not offer an explanation 
for the correlations between the radial velocity amplitude and [Fe/H]
value (Fig.~\ref{fig:AFe}, lower panel).

The metallicity sensitivity of the wavelength dependence of the
photometric amplitudes (Figs.~\ref{fig:mk}-\ref{fig:deltamk}) can be 
a result of the temperature sensitivity of the differential line 
blanketing effect.

Metallicity is only one of the factors that has an influence
on the pulsation amplitude. Another factor is the position of the
Cepheid within the instability region. This aspect has been 
studied by Kanbur \& Ngeow in a series of papers. In their first
paper (\cite{KN04}) the dependence of the photometric
$V$ amplitude on the $V-I$ colour index (corrected for reddening)
was investigated, based on three galactic systems. The sample consisted
of 154, 634, and 391 Cepheids in the Galaxy, LMC, and SMC, respectively.
It is noteworthy that they found linear relationships between the
$A_V$ and $(V-I)_0$ for Cepheids in each galactic sample, and the
slope of the lines shows a systematics according to the average
metallicity of the host galaxy. Obviously, therefore, a part of the 
scatter in Figs.~\ref{fig:AFe}-\ref{fig:feperhfeperh} 
is caused by the colour (temperature) dependence of the pulsation 
amplitude. The separation of the effects of colour and metallicity 
on the pulsation amplitudes is left to a future study.

In addition to the amplitudes, metallicity is expected to
have an influence on the shape of the light curve. These 
correlations exist for RR Lyrae stars (\cite{KZs95}, \cite{JK96}). 
Because pulsation of Cepheids and RR~Lyrae type stars is driven 
by the same mechanism, one expects that the light curve shape is 
sensitive to the atmospheric metal content for Cepheids, too. 
A study of the metallicity sensitivity of Fourier coefficients 
of the phase curves of Galactic Cepheids is in progress.

As an important by-product, we confirmed the separation of 
Galactic classical Cepheids into short- and long-period ones 
at the limiting pulsation period of 10\fd47, especially by 
Fig.~\ref{fig:AFe}. It is noteworthy that the limiting period
of 10\fd47 is close to the resonance centre of 
$2P_{\rm 2OT} = P_{\rm F}$ of the Cepheid pulsation
(where $P_{\rm F}$ is the fundamental period, $P_{\rm 2OT}$ is
the period of the second overtone). This resonance causes 
the famous Hertzsprung progression of the light curve shape of
Cepheids. The value of the resonance centre is, however, not 
definite: from the observational point of view,
the discontinuity of the Fourier phase parameters near 10 d
as derived from the Cepheid light curves is not sharp at all
(see \cite{Setal08}; \cite{Setal10} for LMC and SMC Cepheids, 
respectively).
From the point of view of theory, the resonance centre is 
model-dependent: models calculated by Klapp et~al. (\cite{Ketal85}) 
predict the minimum 
amplitude for both resonant modes between 10\fd1 and 10\fd4,
quite close to the value of the dichotomy determined in Paper~I.
However, the cause of the different behaviour of short- and 
long-period Cepheids remains open. Simon \& Moffett (\cite{SM85})
assume that the long- and short-period Cepheids reach their limit 
cycles in different ways. Figure~\ref{fig:AFe} is another
manifestation of this phenomenon. The situation is, however, even
more complicated because the resonance centre itself is a function 
of the metallicity of the modelled stars (\cite{SK94}).

\section{Conclusion}
\label{sect_concl}

We confirmed that the observed variety in the atmospheric iron 
content of Cepheids contributes to the wide range of the actual
pulsation amplitudes: more metal-deficient Cepheids tend to
pulsate with larger amplitudes. We were able to reveal the 
metallicity dependence of several parameters ($q$, $k$, $m$, 
$\Delta k$, and $\Delta m$) of Galactic Cepheids. In physical terms, 
these relationships mean that the amplitude of the radial velocity 
variations, the ratio of photometric to radial velocity amplitudes, 
and the wavelength dependence of the photometric amplitudes depend 
on the iron abundance of the pulsating stellar atmosphere. In some 
cases, the metallicity sensitivity of the amplitude parameters is 
not the consequence of the differential line blanketing with the
result that stars of higher metallicity are fainter towards shorter
wavelengths.

These correlations, although not strong, can still be used for
deriving the [Fe/H] value of individual Cepheids from photometric
observational data. Improvement of these correlations is
mainly hindered by the current uncertainties in the spectroscopic
abundance determination: individual [Fe/H] values have an error 
of $0.05 - 0.10$. From multicolour photometric data, we derived a 
photometric [Fe/H] value for 21 Galactic long-period Cepheids, one 
of which (GY~Sge) is lacking prior spectroscopic abundance 
determination.

Availability of direct spectroscopic abundance values 
and multicolour light curves of Cepheids in the Magellanic Clouds 
would extend the newly found correlations 
to more negative [Fe/H] values because the interval of atmospheric 
iron content of the Galactic Cepheids (typically $-0.25 < {\rm [Fe/H]}
< +0.25$) is not particularly wide. Quite a few Cepheids more
deficient in iron are known in the outer spiral arm of the Milky Way
system but these Cepheids are so distant that their available
spectroscopic [Fe/H] values are too uncertain for calibration
purposes. Magellanic Cepheids are much more metal-deficient:
the mean [Fe/H] is $-0.34 \pm 0.03$ for the LMC and
$-0.64 \pm 0.04$ for the SMC (Keller \& Wood \cite{KW06}).
The available observational data on the abundance of individual
Magellanic Cepheids are scanty: photometric metallicity values
determined by Harris (\cite{H81}) using the Washington system 
resulted in metallicities of limited accuracy, and spectroscopic
metallicities for a sample consisting of 12-12 Cepheids in each cloud
(\cite{Retal05}).

The finite interval of actual [Fe/H] values of Cepheids situated 
in different parts of the same galaxy is a drawback when attempting 
to decrease the scatter of the $P$-$L$ relationship for Cepheids 
in individual galaxies and in settling the question of the
metallicity sensitivity of the $P$-$L$ relationship. 
The wide variety of iron abundance values of Cepheids in the same host
galaxy is a reason to be cautious when assigning an `average'
metallicity to all Cepheids in a given galaxy. As a rule, the
determination of the heavy element abundance of external galaxies
(except the nearest ones) is based on the chemical properties of their
bright HII regions (\cite{Zetal94}). Metallicity of these regions, 
however, corresponds to youngest population of stars involving 
long-period Cepheids. Short-period Cepheids are somewhat older
(and evolve slower) and may have formed from a less processed
interstellar matter.

Both the effect of metallicity (of non-blanketing origin)
on the amplitudes revealed from observational data and the 
occurrence of the dichotomy at the limit of 10\fd47 (including 
its relation to the $2P_{\rm 2OT} = P_{\rm F}$ resonance phenomenon)
have to be investigated through appropriate theoretical models of 
stellar pulsation. 

\begin{acknowledgements}
This research was supported by the European Space Agency (ESA) and the
Hungarian Space Office via the PECS programme (contract No.\,98090).
The authors are indebted to Drs. Sz. Csizmadia, M.~Kun, N.~Nardetto, 
C.-C.~Ngeow, and the referee for their remarks leading to a better 
presentation of the results.
\end{acknowledgements}

\bibliographystyle{aa}
{}

\end{document}